\documentclass[12pt]{article}
\usepackage{epsfig}
\usepackage{amsfonts}
\usepackage{amsopn}
\usepackage{amsmath}
\usepackage{verbatim,amsthm}

\title{Non-linear realization of Poincar\'e invariance in the graph-representation of extremal hypersurfaces }

\author{
Jens Hoppe \thanks{e-mail:  hoppe@math.kth.se}
\\
Department of Mathematics,\\
Royal Institute of Technology, \\
KTH, 100 44 Stockholm, \\
Sweden
}

\date{}

\begin{document}

\maketitle

\begin{abstract}
In the Born-Infeld 'harmonic gauge' description of M-branes
moving in $\mathbb{R}^{M+1}$ the underlying $M+2$ dimensional
Poincar\'e - invariance gives rise to an interesting system of
conservation laws showing signs of integrability.
\end{abstract}

Varying
\begin{equation}
S[z]=-\int d^Mxdt \sqrt{1-\dot{z}^2+(\vec{\nabla}z)^2} \label{1}
\end{equation}
yields the quasi-linear PDE
\begin{equation}
(1-z^{\alpha}z_{\alpha})\Box  z +z^{\alpha}z^{\beta}z_{\alpha \beta}=0  \label{2}
\end{equation}
for $z(t,\vec{x})$, with $z^{\alpha}:=\eta^{\alpha
\beta}z_{\beta}=\eta^{\alpha\beta}\frac{\partial z}{\partial
x^{\beta}}$, $(\alpha,\beta=0,1,\ldots,M)$,
$\eta^{\alpha\beta}=diag(1,-1,\ldots,-1)$,
$z_{\alpha\beta}=\frac{\partial^2z}{\partial x^{\alpha} \partial
x^{\beta}}$, $\Box z=\eta^{\alpha\beta}z_{\alpha\beta}$.

At first sight, (\ref{2}) appears to be a scalar field equation in
$M+1$ - dimensional Minkowski space, but $M+2$ additional
conservation laws follow when realizing that $z$ is most naturally
interpreted as a (for notational convenience, the last) component of
a $(M+2)$ - vector
\begin{equation}
 x^{\mu}=\left( \begin{array}{c}
                        t      \\
                        \vec{x}   \\
                        z(t,\vec{x})
                              \end{array} \right) \label{3}
\end{equation}
in $\mathbb{R}^{1,M+1}$ (cp. e.g. \cite{1}). The corresponding
world-volume $\mathcal{M}$, described by (\ref{3})/(\ref{2}), can
easily be shown to have vanishing mean curvature. The
$M+2$-dimensional Poincar\'e - symmetry (non-linearly realized, as
$x^{M+1}=z(t,\vec{x})$) is expressed by
\begin{equation}
\partial_{\alpha}\mathcal{H}^{\alpha \mu}=0  \ \ \ \ (\mu=0,1,\ldots,M+1) \label{4}
\end{equation}
\begin{equation}
\partial_{\alpha}(x^{\mu}\mathcal{H}^{\nu \alpha}-x^{\nu}\mathcal{H}^{\mu\alpha})=0 \label{5}
\end{equation}
\begin{equation}
\mathcal{H}^{\alpha \beta}:=\frac{z^{\alpha}z^{\beta}}{\sqrt{1-z^{\gamma}z_{\gamma}}}+\eta^{\alpha\beta}\sqrt{1-z^{\gamma}z_{\gamma}}
\end{equation}
\begin{equation}
\mathcal{H}^{\alpha,M+1}:=\frac{z^{\alpha}}{\sqrt{1-z^{\gamma}z_{\gamma}}}=\mathcal{H}^{M+1,\alpha}
\end{equation}
(generalising to arbitrary co-dimension is straightforward).

Note in particular that

\begin{equation}
z_{\alpha}\mathcal{H}^{\alpha\beta}=\mathcal{H}^{\alpha,M+1}
\end{equation}
Let me make a few comments how one may be able to \emph{use} (\ref{4}) and (\ref{5}).

First of all note the monotonicity of
\begin{equation}
\int x^{\mu}\mathcal{H}^{00}=C^{\mu}+tP^{\mu}
\end{equation}
($C^{\mu}$ and $P^{\mu}=\int \mathcal{H}^{\mu0} $ being constants,
depending only on the initial conditions); and the explicit
(polynomial; here linear) appearance of $t$. 

Secondly, one could hope that the conservation laws, in particular
the $\mu \ne M+1$ part of (\ref{4}) (containing neither $x^{\alpha}$
nor $z$!), thought of as a \emph{first}-\emph{order} system linear
in the derivatives of the $M+1$-vector
\begin{equation}
z^{\alpha}=\left( \begin{array}{c}
                        \dot{z}      \\
                        -\vec{\nabla}z
                              \end{array} \right),
\end{equation}
resp. $u^{\alpha}=\frac{z^{\alpha}}{\sqrt{1-z^{\gamma}z_{\gamma}}}$,
resp. $\mathcal{H}^{\alpha 0}$ (or other suitably chosen coordinates;
e.g. in order to apply the method of characteristics) are
\emph{integrable}.

\vspace{12pt}
\noindent{\bf Acknowledgments}
I would like to thank M. Bordemann, B. Gustafsson, E. Langmann, J. Reinhardt, M. Trzetrzelewski and A. Zheltukhin for valuable discussions.

\end{document}